# When more precision is worse: Do people recognize inadequate scene representations in concept-based explainable AI?

Romy Müller & Wiebke Klausing

Chair of Engineering Psychology and Applied Cognitive Research, TUD Dresden University of Technology

**Abstract**

Explainable artificial intelligence (XAI) aims to help uncover flaws in an AI model's internal representations. But do people draw the right conclusions from its explanations? Specifically, do they recognize an AI's inability to distinguish between relevant and irrelevant features? In the present study, a simulated AI classified images of railway trespassers as dangerous or not. To explain which features it has used, other images from the dataset were shown that activate the AI in a similar way. These concept images varied in three relevant features (i.e., a person's distance to the tracks, direction, and action) and in an irrelevant feature (i.e., scene background). When the AI uses a feature in its decision, this feature is retained in the concept images, otherwise the images randomize over it (e.g., same distance, varied backgrounds). Participants rated the AI more favorably when it retained relevant features. For the irrelevant feature, they did not mind in general, and sometimes even preferred it to be retained. This suggests that people may not recognize it when an AI model relies on irrelevant features to make its decisions.

## Introduction

Explainable artificial intelligence (XAI) aims to help people understand how AI models make decisions. For instance, to explain why a situation depicted in an image was classified as dangerous, it could be shown which image areas the AI has used or what it has seen there. Such XAI is not only promising from a practical perspective, it also brings up exciting questions about human cognition. One interesting question is how people use inconsistencies to infer how other agents (human or artificial) represent categories. For instance, when others use particular features consistently or inconsistently, what does this say about their understanding of a category?

Imagine an AI image classifier decides that a situation is dangerous. To explain why, several other images are shown that activate its neurons in the same way. Probably, you would prefer it if all these images depicted a similar danger source (e.g., person close to railway tracks while a train is approaching). Now imagine that all images also showed the same scene background. Would you evaluate this positively as it indicates that the AI has formed a precise representation of the scene? Or would you conclude that it lacks adequate scene understanding as it uses irrelevant features to decide about danger? This might make you worry whether the AI has formed a sufficiently robust representation of danger, thus being unable to also recognize it in other situations.

A previous study suggests that people want AI models to represent scenes as precisely as possible, without being aware of the potential costs to the AI's robustness (Müller, 2025b). Specifically, participants rated AI performance higher when the concept images used for explanation closely replicated the classified scene than when they generalized over a less relevant feature. In the present study, we examined this preference for precision more systematically. That is, we investigated what image features people do or do not prefer to be retained in concept-based XAI. Before describing the study, we will discuss the problem of AI relying on irrelevant features, the potentials of XAI to reveal such biases, and how humans might respond to these explanations.

## The trade-off between precision and robustness

Obviously, AI models in general and image classifiers in particular should accurately represent the inputs they receive. At the same time, their representations should not be overly precise: they should only include truly relevant features. Yet, representing irrelevant features is a common problem in AI. For one, they can be included *in addition to* relevant features. This is what happens in the case of overfitting (Ying, 2019). It can occur when AI models are trained too extensively and thus develop overly precise representations. A resulting problem is that the AI will no longer make correct decisions when the context changes. Such low robustness of image classifiers due to their inability to generalize is a serious problem (Muttenthaler et al., 2024).

A second problem occurs when the AI uses irrelevant features *instead of* relevant ones, which commonly results from dataset bias (Torralba & Efros, 2011). When relevant and irrelevant features are correlated in a dataset (e.g., all dangerous images were shot in the same location), the AI may not actually learn the relevant features (e.g., person close to tracks) but use the correlated ones as a shortcut. Such shortcut learning is a common phenomenon in image classification (Geirhos et al., 2020). A prevalent bias is that AI models rely on the scene background and thus can no longer classify images when the background changes (Beery et al., 2018). For instance, they fail to recognize cows on a beach, or they falsely recognize sheep when in fact there is nothing but green grass.

While overfitting and shortcut learning have different origins (e.g., too much training vs. dataset bias), a common theme is that the AI relies on irrelevant features. It either uses them in addition to the relevant features (overfitting) or instead of them (shortcut learning). Both can lead to a lack of robustness: the AI will not work in new situations that do not share the same associations between relevant and irrelevant features. Thus, these problems should be detected, and XAI aims to facilitate this detection.

## Can concept-based XAI reveal overly precise AI representations?

What does an AI see in an image? Concept-based XAI can make this transparent by showing the concepts an AI has used to make its decision (for an overview see Poeta et al., 2023). One approach is to show other images or image patches from the dataset that activate the AI is a similar way to the classified image (Achtibat et al., 2023; Fel et al., 2023). Empirical studies have shown that such concept-based XAI can actually help people detect AI biases (Achtibat et al., 2023; Adebayo et al., 2022; Colin et al., 2022; Qi et al., 2021). As a result, they were better able to recognize it when AI models relied on irrelevant features such as the scene background or other data artefacts.

However, in these studies, the XAI typically presented unary or individual features (e.g., watermarks on radiological images). These features can easily be isolated in concept images, focusing the human evaluator's attention on a small image patch which contains all that is needed. This makes it rather easy to recognize which features were used, and also to conclude that the AI should not actually use them. Conversely, the classification of complex categories like danger often relies on non-unary or relational features (e.g., a person's distance to railway tracks) as well as combinations of multiple features (e.g., distance and action). In this case, the concept images necessarily include several features at the same time, which a human evaluator needs to process in parallel. For instance, such concept images might show a person embedded in a scene, rather than an isolated patch of background. If the background contributed to the AI decision, this would be reflected in its repetition: the same background would appear in the classified image and across all the concept images. Presumably, this makes it more difficult to detect AI biases, because they need to be inferred from the repetition of irrelevant features across multiple images. Moreover, people would have to conclude that this repetition is undesirable. In other words, they would have to punish the AI for being overly precise, including an irrelevant feature rather than randomizing over it. The following section will discuss whether this is something we can expect.

## Might people appreciate less precision?

Whether people appreciate less precision in XAI concepts will likely depend on how they interpret it: as a flaw of the AI or as a sign of its ability to generalize. This is not trivial, because humans and AI do not generalize in the same way (Ilievski et al., 2025; Thalmann & Schulz, 2025). Thus, people might not even recognize it when an AI is unable to form sufficiently general representations. Indeed, a previous study suggests that people prefer precise representations over generalizations and do not distinguish the latter from systematic mistakes (Müller, 2025b). For instance, when the classified image showed a person walking on railway tracks, the XAI suggested a more general concept like "being on tracks" by showing images of people engaged in various activities on railway tracks. This made participants rate the AI's performance lower than when the action from the classified image was retained precisely, and just as low as systematic mistakes where all images showed a wrong action (e.g., standing instead of walking). However, one might argue that when evaluating danger, it can be problematic to generalize over actions, so action is not really an irrelevant feature. This raises the question of how the relevance of features affects people's perception of imprecision in XAI. Would they be more positive about an AI's imprecise representations when these representations generalized over completely irrelevant features?

## Present study

The conceptual representations of image classifiers should not include irrelevant features. This also means that concept-based XAI would only retain relevant features and randomize over irrelevant ones. Do people appreciate the resulting imprecision? Do they recognize that some variation is beneficial for the sake of robustness? The present study addressed these questions in a railway safety scenario. Participants saw static images of trespassers in the vicinity of railway tracks and evaluated a simulated AI that had decided whether each situation was dangerous. These decisions were explained by showing other images from the dataset that, according to the AI, were similar in features that informed its decision.

Three of these features were relevant from a human point of view: a person's distance to the tracks, the direction in which the person was turned relative to the tracks, and the person's action. However, their relevance can be expected to vary. First, distance has repeatedly been found to have the strongest impact on human evaluations of AI performance (Müller, 2025a, 2025b). Second, for direction there are no empirical data, yet, but it predicts whether a collision is possible and might thus be considered about as relevant as distance. Third, action is considered a less critical feature, but participants still prefer it when the AI represents actions precisely (Müller, 2025b). Finally, a fourth feature was clearly irrelevant: the scene background. In actual railway traffic, the environment is an important predictor of danger, and train drivers can draw various inferences from it (Müller & Schmidt, 2025). However, naïve participants presumably are unable to do the same. More importantly, the images used here were all taken at a railway testing facility and thus their background only varied in surface features.

The critical experimental manipulation was which features were the same or varied across concept images. We expected an overall sameness preference but hypothesized that it is modulated by feature relevance. That is, participants should more strongly prefer sameness for more relevant features (i.e., distance, direction, action, in that order). However, the most interesting outcome concerned the irrelevant feature (i.e., background). Basically, three outcomes are conceivable.

First, if participants act normatively, their preference should reverse, with higher ratings when the background varies than when it is the same. Given that people seem to prefer precision (Müller, 2025b), we do not consider this a likely outcome. Second, if participants do not care whether an irrelevant feature is retained or not, their ratings should not differ between same and varied backgrounds. Third, if they prefer the AI to represent scenes as precisely as possible, they might even exhibit a sameness preference for the background.

## Methods

### Participants
Fifty-nine participants were recruited from the university's participant pool and received partial course credit or 10 € per hour. Their age varied between 19 and 61 years ($M = 28.1$, $SD = 9.0$), 41 were female and 18 were male.

### Apparatus and stimuli
**Lab setup** Up to four participants worked in parallel in a quiet lab room. The instructions and experiment ran on one of four desktop computers (24") and a standard computer mouse was used as an input device.

**Instruction** An instruction video (5:30 min) was based on a Microsoft PowerPoint presentation with audio overlay. The video introduced the railway scenario and explained the principles of concept-based XAI (i.e., presenting images from the dataset that activate the AI in a similar way to the classified image). Using the toy example of recognizing a dog, participants were informed that concept images could adequately match the classified image, be too dissimilar, too similar, or ill-suited to the decision. Most importantly, they were explicitly informed that too much similarity raises the question of whether the AI will be robust enough to also recognize other images. Moreover, participants were told that for complex concepts like danger, it is not only important what objects are present in the image but what is going on in the scene. Finally, the experimental procedure was explained using screenshots from the experiment.

**Stimuli** All stimuli had a resolution of 1920 × 1080 pixels and consisted of images, text (in German) and interaction elements. Participants saw two screens. The decision screen showed the classified image (650 × 225 pixels), the AI decision (i.e., "dangerous" in red font or "not dangerous" in green font) and two buttons to indicate whether this decision was correct or incorrect, respectively. The AI rating screen (see Figure 1) again showed the classified image and the AI decision but added a set of five vertically arranged concept images (433 × 150 pixels each) as well as a slider to indicate how well the AI has done its job (ranging from "very poorly" to "very well" on a 100-point scale).

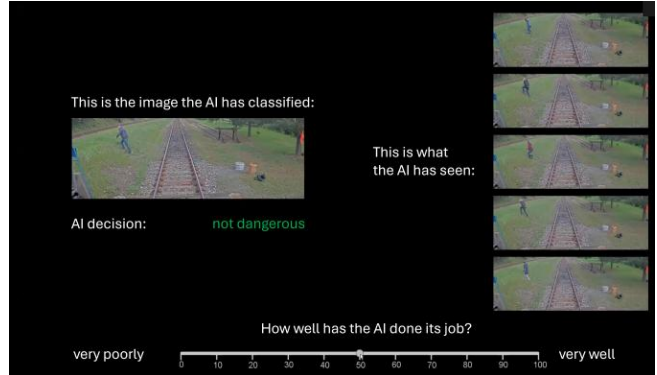
Figure 1: Stimulus example showing an AI rating screen.

There were 128 distinct stimuli, each consisting of a classified image, an AI decision, and five concept images. Between these stimuli, the match between the classified image and the concept images was factorially manipulated in four features: the person's distance to the tracks, the direction in which the person was turned, the action performed by the person, and the scene background. In each feature, the concept images could either be the same as the classified image or they could vary. A feature being the same meant that it corresponded to the classified image in all five concept images (e.g., all images showed people far away from the tracks). Conversely, a feature being varied meant that it corresponded to the classified image in only 2-3 of the concept images but differed in the remaining 2-3 images.

The feature of distance was further subdivided into stimuli where the person was near the tracks versus further away from them (less or more than 1 meter, respectively). Similarly, direction was subdivided into stimuli where the person was turned towards versus away from the tracks, and action was subdivided into stimuli where the person was walking versus performing a different action (e.g., standing, sitting, kneeling). When the action was the same as in the classified image, it was always walking (see Figure 2 for examples).

### Procedure
Before the experiment, participants watched the instruction video. The experiment used a 2 (distance) × 2 (direction) × 2 (action) × 2 (background) within-subjects design. The two levels for each factor were "same" and "varied", respectively. The experiment consisted of 128 trials, resulting from a combination of each of the 64 stimuli with the two AI decisions "dangerous" and "not dangerous". Trial order was randomized individually for each participant. In the first phase of a trial, participants only saw the classified image and AI decision. They indicated whether the AI decision as correct or incorrect by pressing the respective button. This button press brought them to the second part of the trial, removing the buttons and instead revealing the concept images and the slider. On this slider, participants could click any position between 0 and 100 to rate the AI's performance. Taken together, the experiment took between 1 and 1.5 hours.

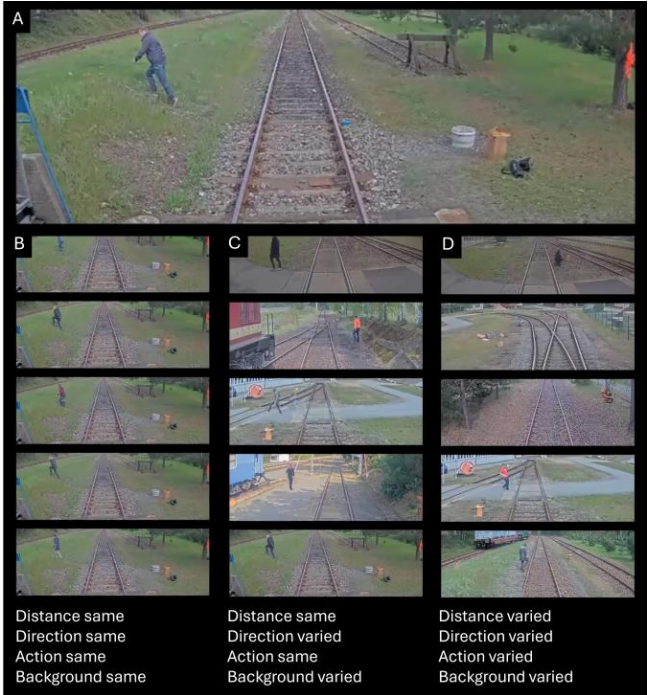

Figure 2: Match between classified image and concepts. (A) Classified image. (B) Same in all features, (C) same in half of the features. (D) Varied in all features.

## Results

### Ratings of AI performance

People's ratings of an AI's performance strongly depend on whether they agree or disagree with its decision (Müller, 2025b). Therefore, agreement was included as a factor in the analysis. The mean ratings of AI performance were analyzed using a 2 (agreement) × 2 (background) × 2 (action) × 2 (direction) × 2 (distance) repeated-measures ANOVA. All pairwise comparisons were performed with Bonferroni correction. For higher-level interactions, we selectively report the effects from the perspective of the irrelevant feature (i.e., background), thus testing whether participants' preference for same versus varied backgrounds depends on the other three features.

For the three relevant features, significant main effects indicated that participants preferred same over varied images (see Figure 3). This was the case for distance (60.1 vs. 48.5), $F(1,58) = 50.436$, $p < .001$, $\eta_p^2 = .465$, direction (58.4 vs. 50.2), $F(1,58) = 39.437$, $p < .001$, $\eta_p^2 = .405$, and action (57.4 vs. 51.2), $F(1,58) = 50.519$, $p < .001$, $\eta_p^2 = .466$. Numerically, the sameness preference was largest for distance, medium for direction, and smallest for action. Conversely, there was no main effect for background, $F(1,58) = 2.800$, $p = .100$, $\eta_p^2 = .046$, indicating that participants' ratings did not differ between same and varied images (55.0 and 53.6).

Moreover, for the two relational features, the sameness benefit was stronger when participants agreed with the AI, as indicated by significant interactions between direction and agreement, $F(1,58) = 19.102$, $p < .001$, $\eta_p^2 = .248$, distance and agreement, $F(1,58) = 4.460$, $p = .039$, $\eta_p^2 = .071$, as well as direction, distance, and agreement $F(1,58) = 5.281$, $p = .025$, $\eta_p^2 = .083$.

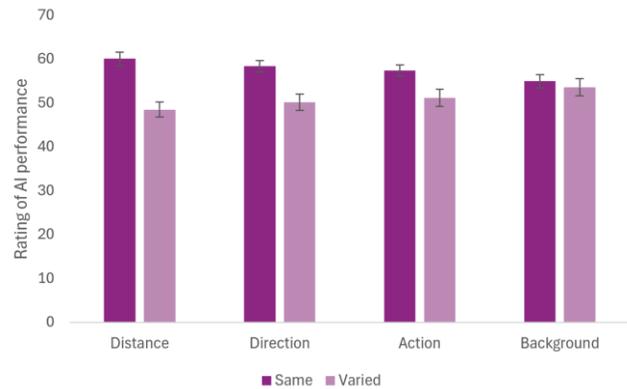

Figure 3: Ratings of AI performance for each feature. Error bars represent standard errors of the mean.

Some of the features also interacted with each other. First, there were interactions between distance and direction, $F(1,58) = 7.420$, $p = .009$, $\eta_p^2 = .113$, as well as direction and action, $F(1,58) = 12.692$, $p < .001$, $\eta_p^2 = .180$. In both cases, the pairwise comparisons reflected a consistent pattern: While the sameness preferences were significant in each comparison, all $ps < .001$, they were larger for a particular feature when the respective other feature also was the same. Moreover, ratings for background depended on distance, $F(1,58) = 6.353$, $p = .014$, $\eta_p^2 = .099$. When the distance was the same, ratings were similar for same and varied backgrounds, $p = .899$, while when the distance varied, ratings were higher for same backgrounds than varied backgrounds, $p = .010$. Thus, participants preferred consistency in the irrelevant feature when the most relevant feature was inconsistent.

In terms of higher-level interactions, a three-way interaction was found between background, action, and distance, $F(1,58) = 4.141$, $p = .046$, $\eta_p^2 = .067$. It indicated that participants preferred the background to be the same when the action was the same but the distance varied, $p < .001$, but were indifferent to the background in all other combinations, all $ps > .2$. Finally, there was a four-way interaction between background, action, direction, and distance, $F(1,58) = 4.438$, $p = .039$, $\eta_p^2 = .071$ (see Figure 4). Pairwise comparisons tested whether there was a preference for same backgrounds in each combination of the other features. This yielded a significant background sameness preference when the distance varied while action and direction were both the same, $p = .043$, or when only action was the same but the two relational features of distance and direction varied, $p < .001$. For all other combinations, it made no difference whether the background was the same or varied, all $ps > .07$. Most notably, when all relevant features were the same, it did not matter whether the background was the same or not, $p = .696$. Taken together, these higher-level interactions suggest that same backgrounds are sometimes preferred, particularly when the most relevant features vary.

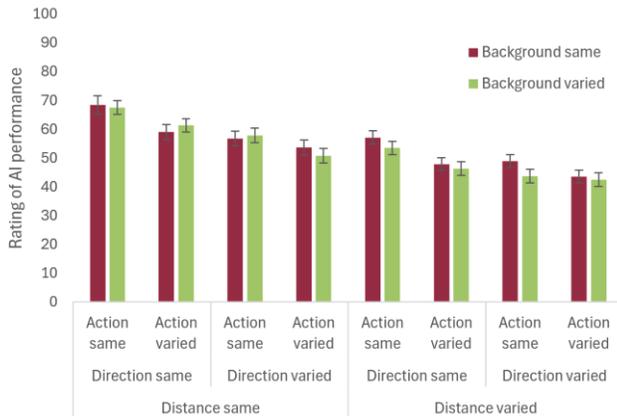

Figure 4: Background sameness effects for all combinations of the other three features. Error bars represent standard errors of the mean.

**Correlations and interindividual differences**

Two exploratory analyses were conducted to gain a better understanding of participants' rating strategies. First, we asked whether the sameness preferences for different features depend on each other. To test this, correlations of the difference between each feature's ratings for same and varied images were calculated. There was a high and significant correlation between direction and action, $r = .605$, $p < .001$. Thus, participants who appreciated images showing the same direction also appreciated images showing the same action. No other correlations between features were significant, all $r$s < .3, all $p$s > .08.

A second exploratory analysis zoomed in on individual participants' ratings for background. For this irrelevant feature, no overall difference had been found between same and varied images. Is this because participants did not care about the background or because some prefer sameness while others prefer variation, so that the opposing preferences average out? To test this, we checked how many participants had a value difference between same and varied images of at least +/- 5 rating units (on a scale from 0 to 100). The vast majority (40 participants, 67.8 %) showed no such difference, while 14 participants (23.7 %) preferred sameness and only 5 participants (8.5 %) preferred variation. Thus, the majority of participants did not seem to care whether the AI included the background in its representations.

## Discussion

Do people always want AI systems to be as precise as possible in their representations, or do they appreciate some imprecision for the sake of higher robustness? In this case, they should prefer it when concept-based XAI explanations vary in irrelevant features. Such variation would speak for the AI's ability to generalize beyond the details of a particular situation. To assess how imprecisions affect the evaluation of AI, participants had to rate AI performance in a railway safety scenario. The concept images used to explain the AI's decisions contained four features with different relevance to the target concept of danger. These features either were the same as in the classified image or they varied across images. Normatively, participants should rate the AI higher when all concept images retain the relevant features (i.e., distance, direction, action) but randomize over the irrelevant feature (i.e., background). However, the results paint a different picture. Participants preferred sameness in general, and the size of this preference increased with increasing feature relevance. Most importantly, they never rewarded variation of the irrelevant feature. This outcome is concerning and will be discussed in the following sections.

**Variation of irrelevant features**

Overall, participants' ratings did not differ between same and varied backgrounds. Does this mean that they did not care whether the AI used the background to decide whether a situation was dangerous? Not necessarily. Another option is that participants might not be answering the right question when evaluating AI performance. That is, instead of rating the AI's internal representations, they might simply rate the relevance of the presented features. If a feature is irrelevant to danger, they simply ignore it. The tendency to exchange a complex question for a simpler one is referred to as substitution bias, which is common in many decision-making contexts (Kahneman & Frederick, 2002). Similar issues have been observed in the context of evaluating the performance of an AI system for person detection (Müller, 2025a). Participants were explicitly informed that the AI was only designed to detect people but was unable to evaluate whether situations were dangerous. Still, they rated its performance lower when it missed people who were in danger. Thus, when people evaluate AI, they might not focus on what the AI is representing and how it is using these representations to perform its assigned task. Instead, they seem to rely on their own prior knowledge about what matters in a scene to make good decisions. An obvious risk of such substitution is that people might not become aware of AI flaws, and thus also not counteract them.

**Interactions between features**

Does people's inadequate use of the scene background depend on whether the other features are represented correctly? For one, this would indicate whether they are insensitive to overfitting or shortcut learning. Moreover, it might help us understand what people are actually doing when they are evaluating AI.

Regarding the first issue, participants seemed indifferent about overfitting and even encourage shortcut learning. If an AI is overfitting, it will use irrelevant features *in addition* to relevant ones. Pairwise comparisons indicated that when all relevant features were correctly represented, participants' ratings did not differ between same and varied backgrounds. Thus, they did not seem to detect overfitting, but also not reward it. Conversely, an AI that has learned shortcuts will use irrelevant features *instead* of relevant ones. In this case, participants even reward the AI's inadequate representations. Specifically, significant preferences for same backgrounds were observed when the distance and direction varied.

What might this tell us about participants' evaluation strategies? One possibility is that they wanted at least something about the AI's representations to match. That is, when the AI got the relevant relational features wrong, it should at least be able to recognize the overall scene. Another possibility is that participants transferred the experience gained from explanations between humans. Humans and XAI differ in how they provide explanations (Qi et al., 2024). For instance, people strategically focus on some aspects in their explanations while leaving out others, and this allows addressees to make inferences about the state of the world (Kirfel et al., 2022). If you want to make a relevant difference salient, it might be best to only vary the feature of interest but keep irrelevant details constant. Indeed, in the present study, participants' ratings differed more strongly between same and varied directions when the backgrounds were consistent. Thus, consistency of irrelevant features might be regarded as an explanatory virtue as it facilitates focusing on relevant differences. However, this is highly speculative, and to date it is far from clear how and why people's preferences for consistency in some features depend on the state of other features.

**Can biases in evaluating AI simply be fixed?**

One might argue that it is easy to fix participants' inadequate evaluation strategies observed in the present study. After all, you just need to tell them explicitly that it is bad when concepts show the same background. However, it seems questionable whether this would fully solve the problem. For one, participants were in fact instructed that concepts could be too similar. This was even illustrated with an example and the risks for an AI's robustness were made explicit. Of course, the instruction could always be adapted to ensure that participants really understand the problem and can transfer it to their own situation.

A more serious concern is that in real life, overspecification can be difficult to detect. In the present study, the consistency of features in the concept images was reduced to a binary variable (i.e., same/varied). For instance, when the AI relied on the background, this resulted in an exact repetition of the background across images, which is easy to detect. In a more realistic setting, it would be much harder to become aware of too much similarity in concept-based XAI. For instance, it could mean that the XAI only shows concept images from rural areas, or only images taken in sunny weather. Thus, the exact scene never repeats, and it is challenging to even notice the undesirable similarities. From this perspective, it is quite concerning that participants do not even seem to mind overspecification when it is as obvious as in the present study.

**Limitations and outlook**

Before drawing conclusions from the present study, some of its limitations should be considered. From a practical point of view, there are several threats to external validity. We only used static images instead of dynamic scenes, simulated instead of real AI and XAI, and novice participants instead of railway safety experts or AI experts. All of these factors are likely to have an impact on the results.

Moreover, there also were limitations of internal validity. One is that we did not go to the limits of sameness. Even in trials where all four features were retained, there still were difference in additional features not considered here, such as a person's identity, clothing, or movement phase. Thus, we cannot actually draw the conclusion that participants prefer the AI's concepts to match the classified image as closely as possible. Presumably, there are limits to their sameness preference, and perhaps even tipping points. Future studies should investigate how hard you need to push the boundaries of similarity in concept images before participants realize that variation is desirable. For instance, the same person could appear in all images, suggesting that the AI relies on person identity. It would be interesting to test whether an even more extreme overspecification would eventually lead to a shift in participants' overall strategies, making them more critical about undesirable repetitions in general.

A more conceptual limitation is that we did not contrast generalizations and systematic mistakes. The varied concepts were generalizations in the sense that the AI randomized over a feature (e.g., varied distances). Conversely, there were no trials in which it used one and the same wrong feature in all images (e.g., all people being far away from the tracks while the person in the classified image is close). Thus, we cannot tell how the type of imprecision (i.e., random or systematic) would affect participants' ratings. A previous study found that for action, it made no difference (Müller, 2025b). However, the present findings suggest that action is perceived as a feature of medium relevance. Thus, future studies should contrast generalizations and systematic mistakes for features of high and low relevance.

**Conclusion**

Do people adequately evaluate the internal representations of AI models when receiving concept-based explanations? That is, do they appreciate beneficial imprecisions that ensure an AI's robustness? The present study suggests otherwise. Participants did not seem to care whether a concept-based XAI suggested that the AI had included the scene background in its representations and decisions. This might indicate that concept-based explanations do not enable participants to draw valid conclusions about the internal representations of AI systems. Thus, an exciting endeavor for future research will be to explore how humans understand an AI's conceptual representations. This might ultimately enable us to better align the perspectives of human and artificial agents.

# Acknowledgments

This work was supported by the German Centre for Rail Traffic Research (DZSF) at the Federal Railway Authority within the project "Explainable AI for Railway Safety Evaluations (XRAISE)"